\documentclass[pra,aps]{revtex4}
\begin{document}

\title{Reply to Yang et al.'s comment}
\author{Nguyen Ba An}
\email{nbaan@netnam.org.vn}
\affiliation{Institute of Physics and Electronics, 10 Dao Tan, Thu Le, Ba Dinh,
Hanoi, Vietnam}

\begin{abstract}
This is to reply to a recent comment by Yang, Yuan and Zhang on ``Teleportation 
of two-quNit entanglement: Exploiting local resorces''.
\end{abstract}


\maketitle
The authors of the recent comment \cite{r1} have taken their patience to recheck the calculation in \cite{r2} 
on ``Teleportation of two-quNit entanglement: Exploiting local resorces''. 
However, Method 1 in \cite{r1} would not be necessary. 

The simplest way to make all the formulae in \cite{r2} remain valid is just to do 
the exchange of indices $3\leftrightarrow 4$ in the whole text of \cite{r2}. 

Or, as also noticed in \cite{r1} (Method 2), one needs just replace $X$ ($Y$) by 
$Y$ ($X$) in the definition (9) in \cite{r2}, without making the exchange  
$3\leftrightarrow 4$. In the latter option, the states $|\Phi_{mn}\rangle_{13}$, 
with $m=1,2$ and $n=0,1,2$, above Eq. (2) 
in \cite{r2} should be changed to
$$
|\Phi_{10}\rangle_{13}=\frac{1}{\sqrt{3}}(e^{-\frac{2\pi i}{3}}|20\rangle+e^{\frac{2\pi i}{3}}|11\rangle +|02\rangle)_{13},
$$
$$
|\Phi_{11}\rangle_{13}=\frac{1}{\sqrt{3}}(e^{\frac{2\pi i}{3}}|10\rangle+|01\rangle+e^{-\frac{2\pi i}{3}}|22\rangle)_{13},
$$
$$
|\Phi_{12}\rangle_{13}=\frac{1}{\sqrt{3}}(|00\rangle+e^{-\frac{2\pi i}{3}}|21\rangle+e^{\frac{2\pi i}{3}}|12\rangle)_{13},
$$
$$
|\Phi_{20}\rangle_{13}=\frac{1}{\sqrt{3}}(e^{\frac{2\pi i}{3}}|20\rangle+e^{-\frac{2\pi i}{3}}|11\rangle+|02\rangle)_{13},
$$
$$
|\Phi_{21}\rangle_{13}=\frac{1}{\sqrt{3}}(e^{-\frac{2\pi i}{3}}|10\rangle+|01\rangle+e^{\frac{2\pi i}{3}}|22\rangle)_{13},
$$
$$
|\Phi_{22}\rangle_{13}=\frac{1}{\sqrt{3}}(|00\rangle+e^{\frac{2\pi i}{3}}|21\rangle+e^{-\frac{2\pi i}{3}}|12\rangle)_{13}.
$$

This Reply, thanks to \cite{r1}, could also serve as an Erratum to \cite{r2}.

\end{document}